# Tensegrity-Inspired Polymer Films: Progressive Bending Stiffness through Multipolymeric Patterning

*Rikima Kuwada*[*,1,3], *Shuto Ito*[3], *Yuta Shimoda*[2,3], *Haruka Fukunishi*[1,3], *Ryota Onishi*[1,3], *Daisuke Ishii*[1], *Mikihiro Hayashi*[1,3]

[1]Department of Life Science and Applied Chemistry, Graduated School of Engineering, Nagoya Institute of Technology, Gokiso-cho Showa-ku Nagoya-city Aichi Japan, 466-8555

[2]Jun Sato Structural Engineers Co., Ltd., Minato-ku Higashiazabu Tokyo Japan 106-0044

[3]Biomatter Lab, Sakurano-cho Toyonaka-city Osaka Japan, 560-0054

**Running Head**

Bend-stiffening film by membrane tensegrity

**ABSTRACT**

Materials with J-shaped stress-strain behavior under uniaxial stretching, where strength increases as deformation progresses, have been developed through various materials designs. On the other hand, polymer materials that progressively stiffen under bending remain unrealized. To address this gap, this study drew inspiration from membrane tensegrity structures, which achieve structural stability by balancing compressive forces in rods and tensile forces in membrane. Notably, some of these structures exhibit increased stiffness under bending. Using a multipolymer patterning technique, we developed a polymer film exhibiting membrane tensegrity-like properties that stiffens under bending. This effect results from membrane tension generated by rod protrusions and an increase in second moment of area at regions with maximum curvature.

Introduction

Natural materials such as skin[1], ligaments[2], tendons[3], arteries[4], and spider silk[5,6] exhibit J-shaped stress-strain behavior, stiffening rapidly under large deformations. This behavior effectively limits further stretching and prevents structural failure. Inspired by this property, many polymer materials have been designed to stiffen with increasing deformation[7–11]. However, such cases have been limited to uniaxial loading. To our knowledge, no studies have yet reported polymer materials that stiffen under off-axis or out-of-plane deformations, such as bending.

To address this gap, we drew inspiration from membrane tensegrity structures, which balance compressive forces in rods with tensile forces in membrane and have gained significant attention in fields like architecture and structural engineering[12–14]. Notably, some of these structures are designed to increase stiffness with bending(Fig. 1a,b)[13,14]. Inspired by this feature, we created a polymer film with macroscale rods and membranes of differing elastic moduli, aiming to develop a material that stiffens as bending progresses. These materials may have valuable applications across a range of fields. In advanced applications like soft robotics, they can enhance grip strength for object handling by adding out-of-plane stiffness to membranes. In more everyday contexts, they also hold promise for use in sports and rehabilitation braces, where tailoring support strength to specific ranges of motion could improve both safety and effectiveness in injury prevention and recovery.

For this study, we employed a practical approach for integrating and patterning polymers with distinct properties, previously reported by our group[15]. This method utilizes two photocurable monomer solutions and a commercial liquid crystal display (LCD) printer for polymer patterning. Initially, a parent cross-linked film was created using one monomer solution, which was then swollen with the second solution. Area-selective UV irradiation was applied for patterning, with the LCD printer offering precise spatial control of UV exposure through digital processing. This versatile technique enables a wide range of applications by allowing various combinations of monomer types and irradiation patterns. Using this technique, we created samples with and without membrane tensegrity pattern by controlling the rod and membrane arrangement. Two polymers with different elastic moduli were employed to establish a stiffness contrast between the rods and the membrane.

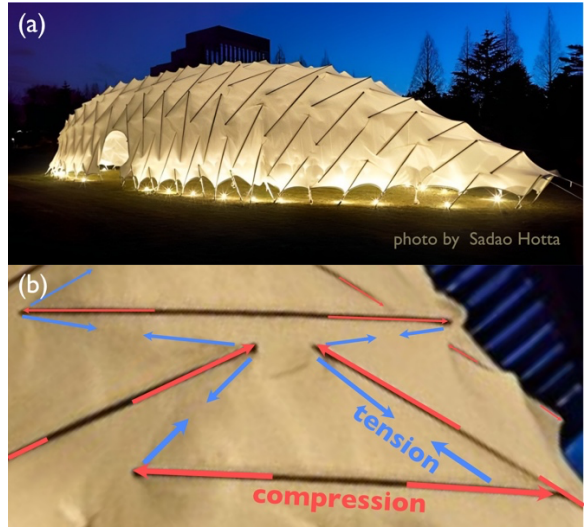

Figure 1. (a) Image of a membrane tensegrity structure that served as the inspiration for the design of polymer films in this study. (b) Image of a membrane tensegrity structure, showing the balance of compressive (red arrows) and tensile forces (blue arrows) within the structure, with rods protruding and the membrane under tension. films in this study.

Using samples fabricated through the above method, we conducted experiments to evaluate their stiffening behavior under additional out-of-plane bending deformations. Given that films are frequently used in three-dimensional configurations in practical applications, it is essential to observe their deformation behavior and load response when further deformations or loads are applied. Accordingly, we first shaped samples with and without a membrane tensegrity pattern into tunnel forms, then applied additional deformations and compared their resulting deformation behavior and load response. Additionally, to explain the deformation behavior and load response of the tunnel-shaped film, we investigated the role of the rod's unique repeating unit patterns within the film. To confirm the progressive stiffening characteristics of these unit patterns, we analyzed their bending moment and bending stiffness.

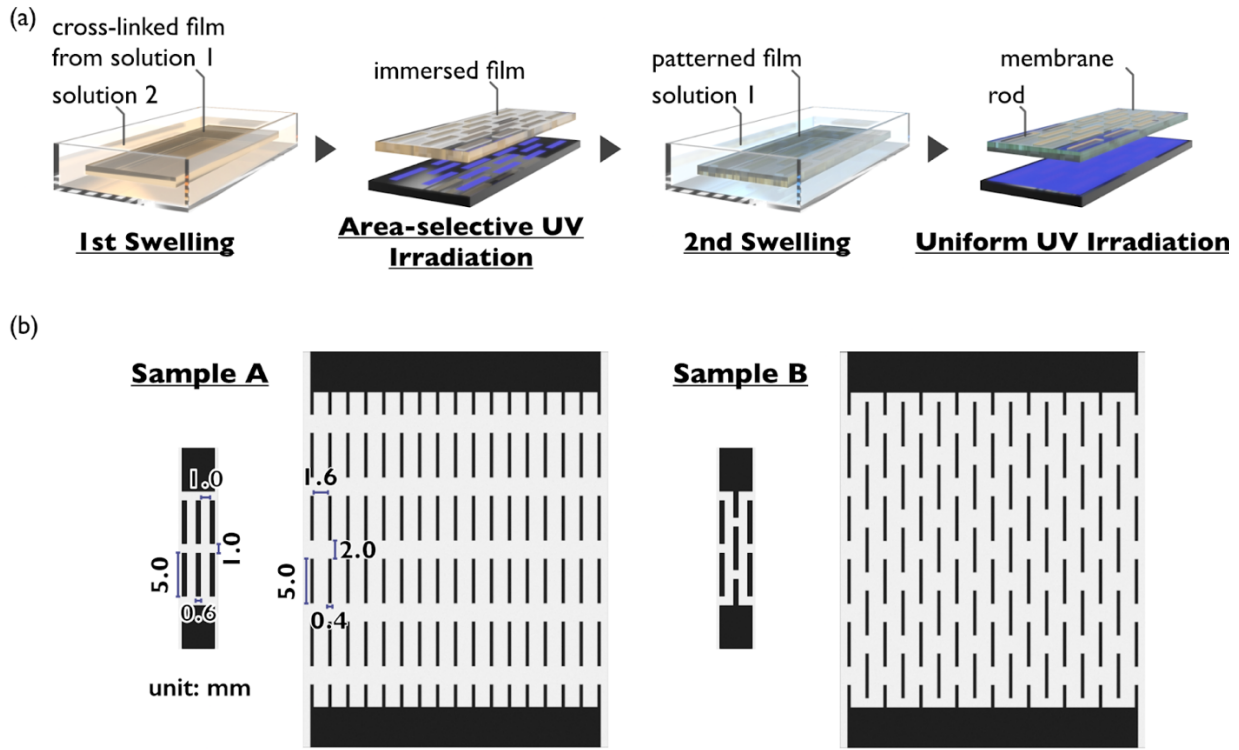

Figure 2. (a) Schematic of the patterning and post-treatment process of samples. (b) Design schematics of samples A and B. The sample with 17 rows of rods is used for load-induced deformation observations and loading-unloading tests, while the sample with 3 rows of rods is used for bending tests.

# Materials and methods

We prepared samples according to the procedure shown in Figure 2a. Solution 1, used to create a flexible membrane, primarily contains methyl acrylate with a small amount of divinyl cross-linker. For rigid rods, we used Solution 2, mainly composed of N,N-dimethylacrylamide with a high proportion of divinyl cross-linker. We conducted tensile tests, load-induced deformation observations, loading-unloading tests, and bending tests. For all tests except the tensile test, two sample types were used: sample A, with horizontally aligned rods, and sample B, with vertically oriented rows of rods, half-phase shifted in alternating rows (Fig. 2b). Details on sample preparation and design for all tests are provided in the supplementary information. In the supplementary information, actual images of the fabricated films (Fig. S1) and details on sample preparation for all tests are provided.

Tensile Test

We evaluated the elastic modulus of the rod and membrane sections, as well as the fracture strain of the membrane, using dogbone specimens made from the rod and membrane materials, respectively. Tensile tests were conducted at a rate of 10 mm/min with a universal testing machine (Autograph AGS-500NX, Shimadzu, Japan).

Load-Induced Deformation Observations

To compare deformation behavior, samples A and B were curved into a tunnel shape, mounted on a jig, and secured in the lower chuck of a universal testing machine. A 10 g weight was applied to the tunnel, and its deformation was recorded.

Loading-Unloading Tests

We compared the deformation behavior and load response of samples A and B under additional loads in three-dimensional shapes. Following the load-induced deformation observations, we mounted the tunnel-shaped sample in a jig and secured it to the lower chuck of the universal testing machine. A plate component was attached to the upper chuck, lowered at 10 mm/min to compress the sample by 7 mm, and then raised at the same speed until the load returned to zero. The load during this stroke was recorded, and maximum load values for samples A and B were compared using Welch's t-test[16].

Bending Tests

We investigated the role of the rod's unique repeating unit patterns within the film. To confirm the progressive stiffening characteristics of these unit patterns, we analyzed their bending moment and bending stiffness. Both ends of each sample were secured in the chucks of a universal testing machine, and bending was applied by moving the chucks 5.5 mm closer at a speed of 0.1 mm/sec. To record the test, a camera was lowered at half the stroke speed using a dip coater (Fig. S2). We calculated the eccentric distance ($e$) and curvature ($\kappa$) of the maximum-curvature region, defined as the area with the greatest curvature during bending, from binary images of each video frame. The sample edges were marked in black to facilitate this measurement (Movie S1,2). Video analysis was conducted in Python 3.10.9, and noise reduction was achieved by applying a Gaussian filter to smooth the eccentric distance and curvature data (Fig. S3,4). Load ($p$) and displacement data were recorded throughout the bending test, and the bending moment ($M$) and bending stiffness ($K_b$) were calculated using equations (1) through (3):

$$M = ep \qquad (1)$$
$$M = K_b \kappa \qquad (2)$$
$$K_b = EI \qquad (3)$$

where $K_b$, $E$, and $I$ represent the bending stiffness, Young's modulus, and second moment of area, respectively. These calculations allowed us to assess the effects of different patterning approaches.

Results and discussion

Tensile tests quantitatively confirmed a significant difference in material properties between the rod and membrane. The elastic moduli of the rod and membrane materials were 809 ± 107 MPa (n = 11) and 1.35 ± 0.19 MPa (n = 14), respectively, indicating an approximately 600-fold difference. The membrane material exhibited a fracture strain of 89.9 ± 9.1% (n = 14), demonstrating sufficient flexibility for bending deformation. For these measurements, both samples had nearly identical thicknesses and widths: the thickness was 0.83 ± 0.04 mm and the width 4.91 ± 0.13 mm in the dogbone specimen made from the rod material (n = 11), and similarly, the thickness was 0.88 ± 0.06 mm and the width 4.07 ± 0.27 mm in the specimen made from the membrane material (n = 14).

Using samples patterned with these distinct materials, we conducted load-induced deformation observations, loading-unloading tests, and bending tests. In load-induced deformation observations, tunnel-shaped samples A and B showed distinct deformation behaviors under a 10 g weight. In sample A, the tunnel shape collapsed due to bending in the rod-free horizontal regions (Fig. 3a). By contrast, sample B maintained its tunnel shape, likely due to a balance of compressive forces in the rods and tensile forces in the membrane. This balance was achieved through slight protrusions of rods from the membrane, generating the necessary compressive and tensile forces (Fig. 3b, Movie S3,4). This behavior suggests the key characteristics of membrane tensegrity.

For quantitative evaluation, we conducted loading-unloading tests on tunnel-shaped samples A and B to compare their deformation behaviors and load response characteristics.

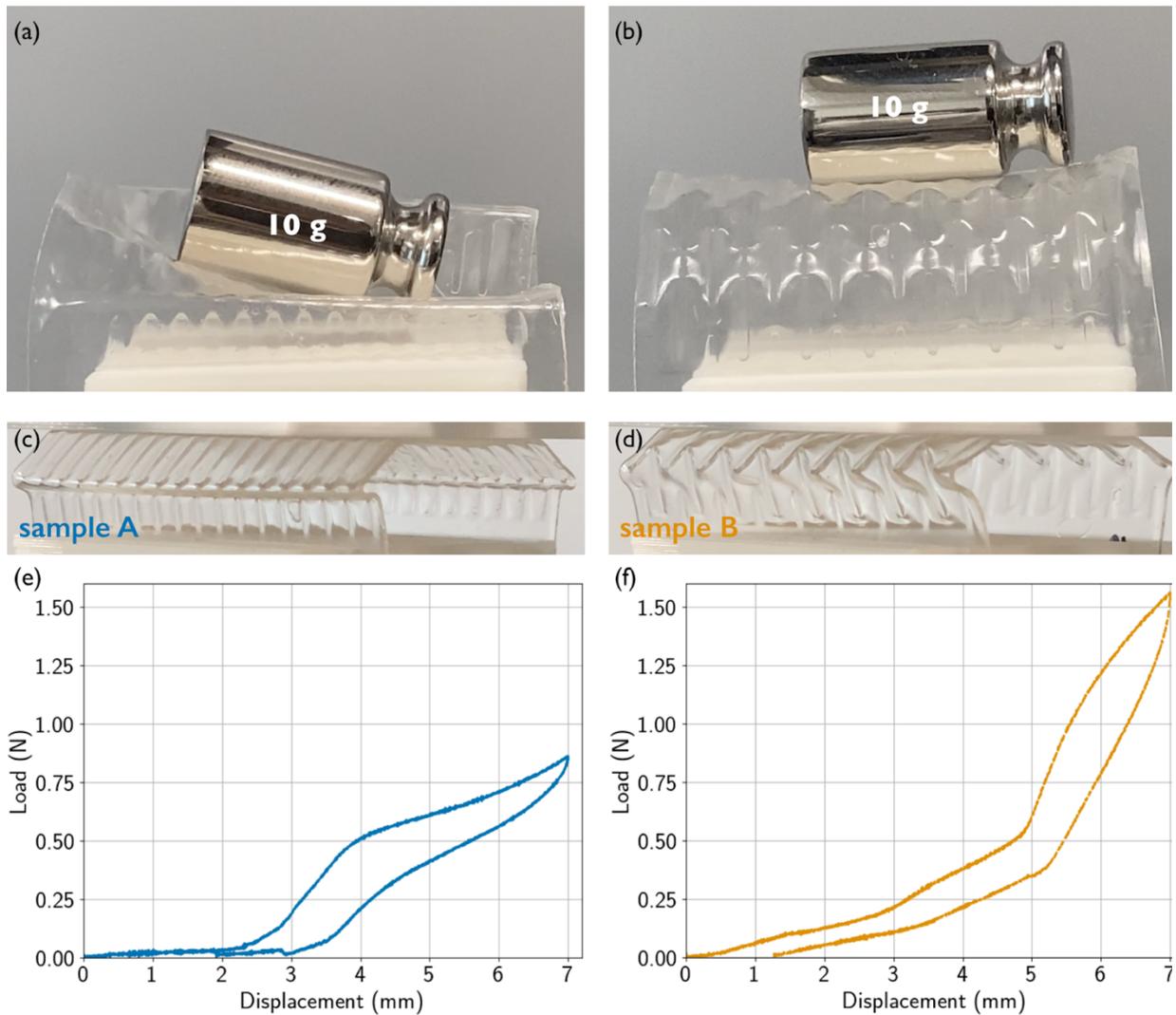

Figure 3. Images of sample A (a) and sample B (b) at a displacement of 7 mm during the loading-unloading test, and load-displacement curves for sample A (c) and sample B (d).

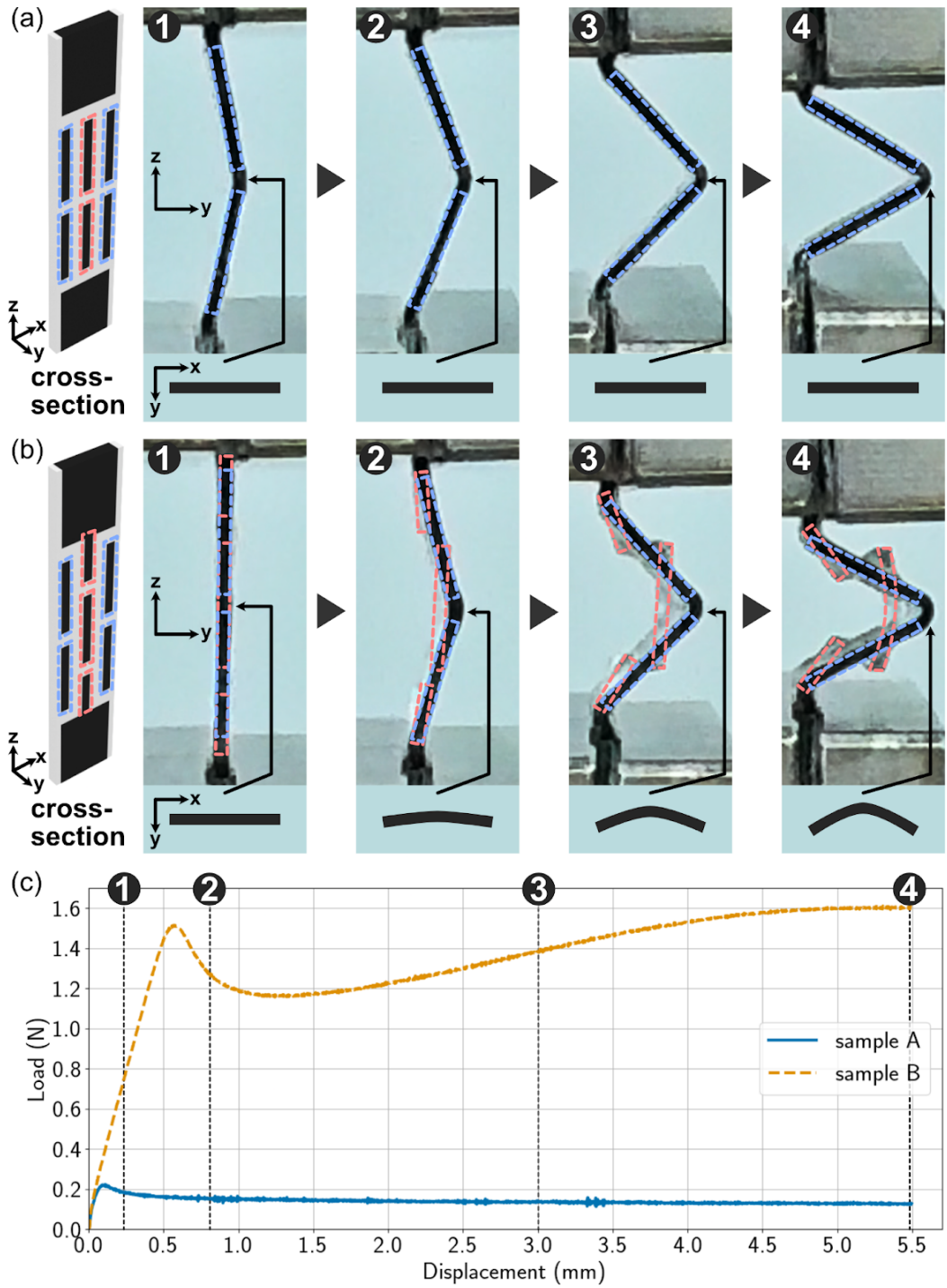

Figure 4. Images of sample A (a) and sample B (b) at displacements of 0.3 mm (①), 0.8 mm (②), 3.0 mm (③), and 5.5 mm (④) during the bending test, along with schematic cross-sectional views of the maximum-curvature region and load-displacement curves for sample A and sample B (c).

Samples A and B exhibited distinct deformation behaviors under progressive displacement, showing differences in load response and structural changes. In sample A, as displacement increased from 0 to approximately 2.5 mm, the tunnel apex expanded and flattened with minimal load increase. Between approximately 2.5 and 4 mm displacement, the tunnel sides began to bulge outward, resulting in a load increase. Subsequently, the rod-free regions along the horizontal axis began bending, which moderated the load increase. During unloading, bending in these regions gradually reversed, restoring the tunnel shape (Fig. 3c,e, Movie S5). In contrast, sample B displayed a different response. As displacement increased from 0 mm to approximately 5 mm, the tunnel apex expanded and flattened, while rods near the apex and sides began to protrude slightly from the membrane. Subsequently, the plate component contacted the side rods, pushing them outward and significantly increasing the load due to heightened membrane tension. Sample B reached a maximum load of 1.56 ± 0.47 N (n = 8), nearly double that of sample A at 0.81 ± 0.16 N (n = 11) (Fig. S5). During unloading, the protruding rods in sample B gradually returned to their original positions, restoring the tunnel shape (Fig. 3d,f, Movie S6). Both samples appear to deform by accumulating elastic strain during loading and recover by releasing this strain during unloading. Figure S6 presents load-displacement curve results from 11 samples of sample A and 8 samples of sample B, confirming reproducibility. In loading-unloading tests, the thicknesses of both samples were nearly identical: sample A measured 0.36 ± 0.03 mm, and sample B measured 0.35 ± 0.05 mm.

To explain the deformation behavior and load response of the tunnel-shaped film, we conducted bending tests to investigate the role of the rod's unique repeating unit patterns within the film. Bending tests revealed significant differences in load progression and deformation between samples A and B. In sample A (Movie S1), bending in the maximum-curvature region caused a slight initial load increase, after which the load gradually decreased and stabilized (Fig. 4a, c). By contrast, sample B (Movie S2) exhibited distinctive structural changes in the maximum-curvature region and a notable load response. Specifically, under vertical compression at a displacement of approximately 0.5 mm, a substantial load increase was observed (Fig. 4b, c①). This increase likely resulted from adjacent rod tips moving apart, stretching the membrane and generating a restoring force (Fig. S7). Unlike sample A, eccentric distance and curvature increased only slightly during this phase. Afterward, both samples displayed a similar monotonic increase in eccentric distance and curvature (Fig. S8,9). With further bending of sample B, the load peaked and then decreased (Fig. 4b, c②). Between approximately 1.5 and 4 mm displacement, with curvature ranging from approximately 1000 to 2000 $m^{-1}$, the rods began to protrude, stretching the membrane and gradually increasing the load due to enhanced membrane tension (Fig. 4b, c③; Fig. S10,11). Afterward, as the central rod began to buckle, the load increase halted. This likely occurred because the rods could no longer sustain the compressive force from membrane tension, which also reduced the reactive force (Fig. 4b, c④). Figures S8 to S12 show bending test results from five samples each of sample A and sample B, confirming reproducibility. Both samples had nearly identical thicknesses, with sample A at 0.81 ± 0.05 mm (n = 5) and sample B at 0.84 ± 0.05 mm (n = 5).

Throughout the bending test, sample B exhibited a more dynamic profile in bending moment and stiffness than sample A, showing multiple changes in response to varying curvature. Sample B consistently showed higher bending moment and stiffness than sample A, indicating greater resistance to bending. According to equations 1 and 2, the slope of the moment–curvature curve represents bending stiffness. In sample A, the slope remained constant, while in sample B, it changed several times (Fig. 5a, Fig. S12). This trend in the moment–curvature curve becomes more pronounced in the stiffness–curvature curve. In sample A, bending stiffness initially showed a slight increase before gradually decreasing as the curvature increased. In contrast, sample B's bending stiffness rose significantly at the start, then decreased, gradually increased again within a curvature range of approximately 1000 to 2000 $m^{-1}$, and finally decreased once more (Fig. 5b, Fig. S13). This pattern suggests that bending stiffness in sample B responds dynamically to load variations influenced by the sample's shape.

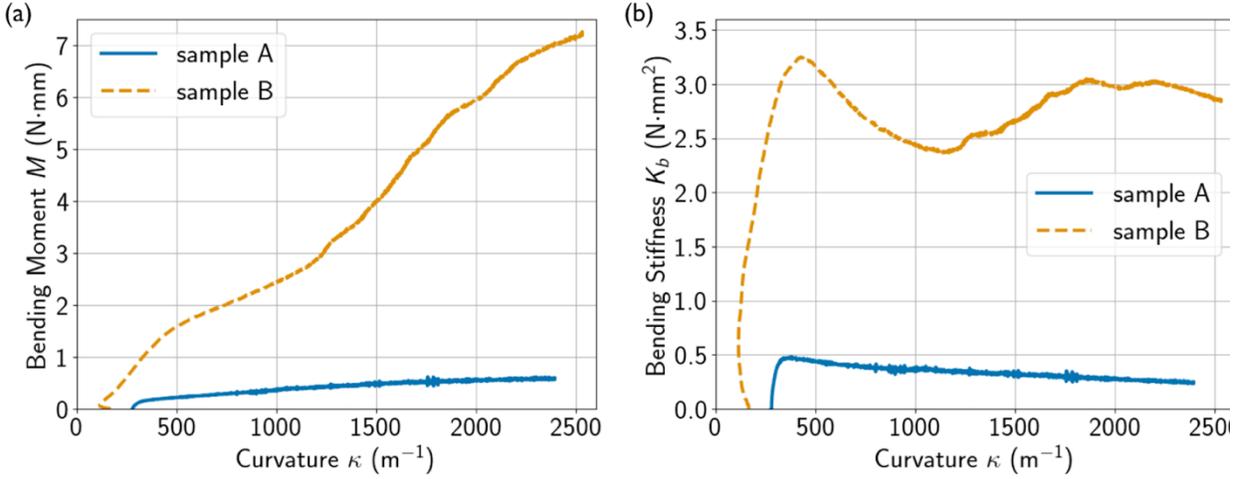

Figure 5. (a) Moment–curvature curves and (b) stiffness –curvature curves for samples A and B.

These differences in bending stiffness between samples A and B appear to stem from distinct structural changes in the maximum-curvature region. In sample A, this region maintained a constant cross-sectional shape during bending, resulting in a single-curved surface (Fig. 4a). In contrast, sample B's cross-sectional shape curved in the opposite direction of bending, creating a saddle surface (Fig. 4b). This saddle shape likely arises from differential deformation, with the membrane surrounding the central row of rods stretching and expanding, while the rods themselves undergo minimal deformation. Consequently, the saddle shape enables efficient stress distribution across the structure by positioning stiff rods in compressed regions and flexible membranes in tensioned areas. As deformation progresses in sample B's maximum-curvature region, the second moment of area increases, further enhancing bending stiffness. These observations suggest a strong correlation between the second moment of area and bending stiffness within the curvature range (approximately 1000 to 2000 m⁻¹) where the rods avoid buckling.

In summary, our observations suggest that the membrane tensegrity pattern developed in this study generates two types of localized deformations under bending. First, the rods protrude, and the membrane stretches; that is, off-axis and out-of-plane deformations are converted into localized uniaxial stretching. Second, the shape in the maximum-curvature region deforms, leading to changes in the second moment of area. Together, these effects result in a gradual increase in bending stiffness.

In this way, our material design concept enables stiffening in response to off-axis or out-of-plane deformations by controlling localized deformation through a macroscopic arrangement of materials with varying elastic moduli. This approach offers valuable insights into achieving J-shaped stiffness responses in these deformation modes. Furthermore, we anticipate that increasing rod stiffness and using materials with J-shaped stress-strain behavior in the membrane could enable true J-shaped bending stiffness responses.

Additionally, this technique is highly versatile, supporting a wide range of polymer compositions and patterning configurations. Researchers can adapt this flexibility to create specific mechanical responses tailored to various applications. For instance, adjusting polymer types and pattern geometries enables precise control over stiffness, flexibility, and deformation behavior, allowing for fine-tuning to meet specific performance needs. This adaptability provides a strong foundation for developing customized solutions in soft robotics, especially for robotic grippers that require controlled stiffness. Adaptive supports and other devices that benefit from dynamic, out-of-plane stiffness could also be optimized through this approach. Future studies could build on these results by exploring new material combinations and intricate geometric patterns to further enhance and refine mechanical responses.

Conclusion

This study demonstrated the feasibility of designing polymer films that progressively stiffen under bending. Such progressive stiffening has previously been limited to uniaxial deformation modes. Inspired by membrane tensegrity structures and using a multipolymer patterning technique, we successfully fabricated a polymer film featuring macroscale rods and membranes with differing elastic moduli. This design enabled the material to resist bending through two key mechanisms: tensile forces generated as rods protrude from the membrane and an increase in the second moment of area within maximum-curvature regions. These findings suggest that such materials could be key to achieving J-shaped stiffness responses under non-uniaxial deformation modes. This approach offers a promising foundation for broader functional material design. Additionally, the versatility of this technique allows for various polymer compositions and patterning configurations. This flexibility provides a platform to tailor mechanical responses for specific applications. Future work could build on these findings to explore diverse material combinations and geometric patterns. Such exploration would aim to fine-tune responses for applications like soft robotic grippers, adaptive supports, and other devices where dynamic, out-of-plane stiffness is advantageous.

ACKNOWLEDGMENT

This work was supported by JSPS KAKENHI Grant Number JP23K23917 (D.I.). and Integrated Research Project of NITech Frontier Research Institutes in Nagoya Institute of Technology

AUTHOR INFORMATION

Corresponding Author

*E-mail: rikima.kuwada@gmail.com

Author contributions

R.K.,S.I., and Y.S. conceptualized the study. R.K.,S.I.,Y.S.,H.F., and R.O. designed the experiments. R.K. collected data. R.K.,S.I., and Y.S .analyzed the data. S.I.,Y.S., and M.H supervised the study. R.K. wrote the first draft. R.K.,S.I., and M.H. edited the manuscript. M.H and D.I. provided equipment and financial support. All authors discussed the results and gave the final approval for publication.

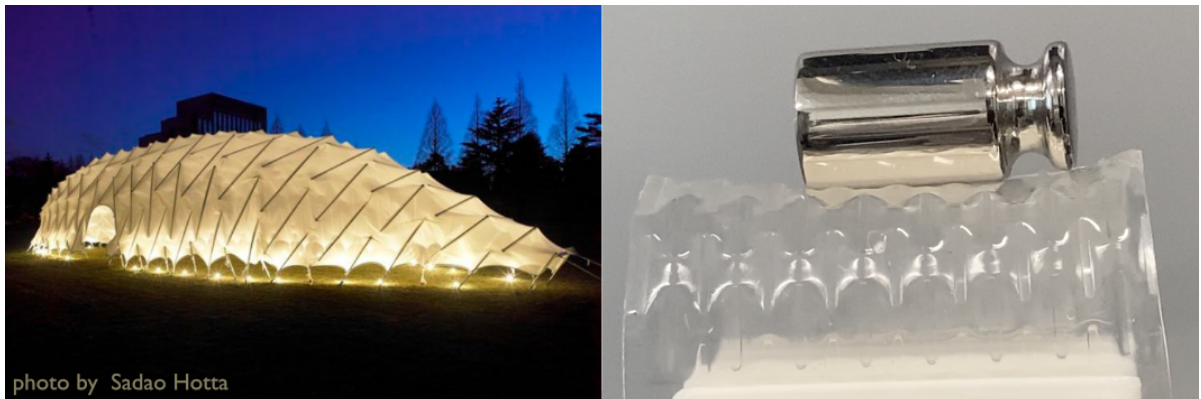

graphical abstract

**Supplementary information**

**Tensegrity-Inspired Polymer Films: Progressive Bending Stiffness through Multipolymeric Patterning**

*Rikima Kuwada*[*,1,3], *Shuto Ito*[3], *Yuta Shimoda*[2,3], *Haruka Fukunishi*[1,3], *Ryota Onishi*[1,3], *Daisuke Ishii*[1], *Mikihiro Hayashi*[1,3]

[1]Department of Life Science and Applied Chemistry, Graduated School of Engineering, Nagoya Institute of Technology, Gokiso-cho Showa-ku Nagoya-city Aichi Japan, 466-8555

[2]Jun Sato Structural Engineers Co., Ltd., Minato-ku Higashiazabu Tokyo Japan 106-0044

[3]Biomatter Lab, Sakurano-cho Toyonaka-city Osaka Japan, 560-0054

Corresponding author: Rikima Kuwada (rikima.kuwada@gmail.com)

**List of Supplementary Information**

Sample preparation and design

All reagents used in this study were purchased from Tokyo Chemical Industry Co., Ltd. (TCI). The samples used in the experiments were prepared according to the procedure shown in Figure 1. Sample preparation involved several polymerization reactions, including the fabrication of the parent cross-linked film. Each reaction was conducted via free radical polymerization. Solution 1 was used for the fabrication of the parent cross-linked film and for the second swelling, while Solution 2 was used for the first swelling. Solution 1 contained methyl acrylate (MA), a divinyl cross-linker (1,4-butanediol diacrylate, BDA), and a photo radical initiator (diphenyl(2,4,6-trimethylbenzoyl)phosphine oxide, TPO). The feed ratio was set to [MA]:[BDA] = 1000:1, with TPO at a weight fraction of 1%. Solution 2 contained N,N-dimethylacrylamide (DMAm), BDA, and TPO, with a feed ratio of [DMAm]:[BDA] = 2:1 and a TPO weight fraction of 1%. UV irradiation was performed using a commercial LCD (Liquid Crystal Display) 3D printer. To create a spacer, a polytetrafluoroethylene (PTFE) sheet with a thickness of either 0.1 mm or 0.3 mm was fixed to the printer bed with double-sided tape. Solution 1 was poured into the resin tray of the printer. The entire surface was irradiated with 405 nm UV light for 3 minutes, producing a uniform parent cross-linked film. This film was then swollen with Solution 2 and irradiated through an LCD-projected pattern for 30 seconds, forming rigid rods in the irradiated areas. After patterning, the sample was immersed in tetrahydrofuran (THF). Solvent exchange was conducted by replacing THF four times. The film was then vacuum-treated at room temperature for over 30 minutes to remove residual solvent. In a final step, the film was swollen once more with Solution 1 and uniformly irradiated with UV light for 3 minutes.

This study included tensile tests, load-induced deformation observations, loading-unloading tests, and bending tests. For tensile tests, the parent cross-linked film was first swollen with Solution 2, then cut into shape using a dogbone cutting die. For the first UV irradiation, two types of samples were prepared: one was uniformly irradiated with UV light for 30 seconds, and the other was not exposed. Both samples were then subjected to a second swelling with Solution 1, followed by uniform UV irradiation.

For the load-induced deformation observations, loading-unloading tests, and bending tests, samples were based on two patterns. Sample A had horizontally aligned rods, while Sample B had vertically oriented rows of rods, half-phase shifted in alternating rows. In the loading-unloading test sample, rods were arranged in 5 rows and 17 columns, with each rod measuring 5.0 mm in length and 0.4 mm in width. Horizontal spacing was 1.6 mm, and vertical spacing was 2.0 mm. For the bending test, the rods were arranged in 2 rows and 3 columns, each measuring 5.0 mm in length and 0.6 mm in width, with both horizontal and vertical spacings of 1.0 mm. The rods at the edges of the sample were fused with spacer sections to enable attachment to fixtures.

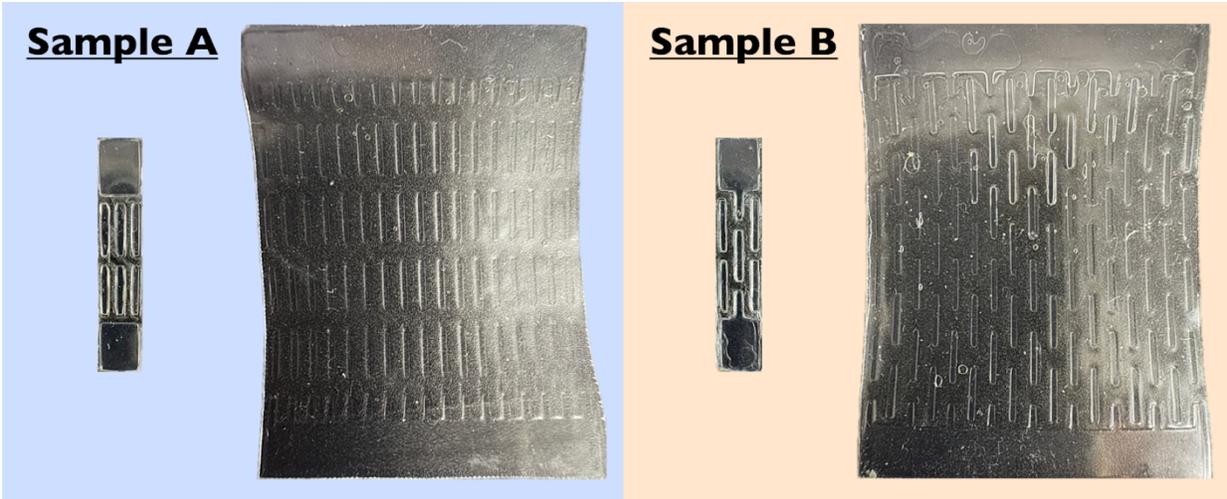

Figure S1. Actual image of sample A and B. The sample with 17 rows of rods is used for load-induced deformation observations and loading-unloading tests, while the sample with 3 rows of rods is used for bending tests.

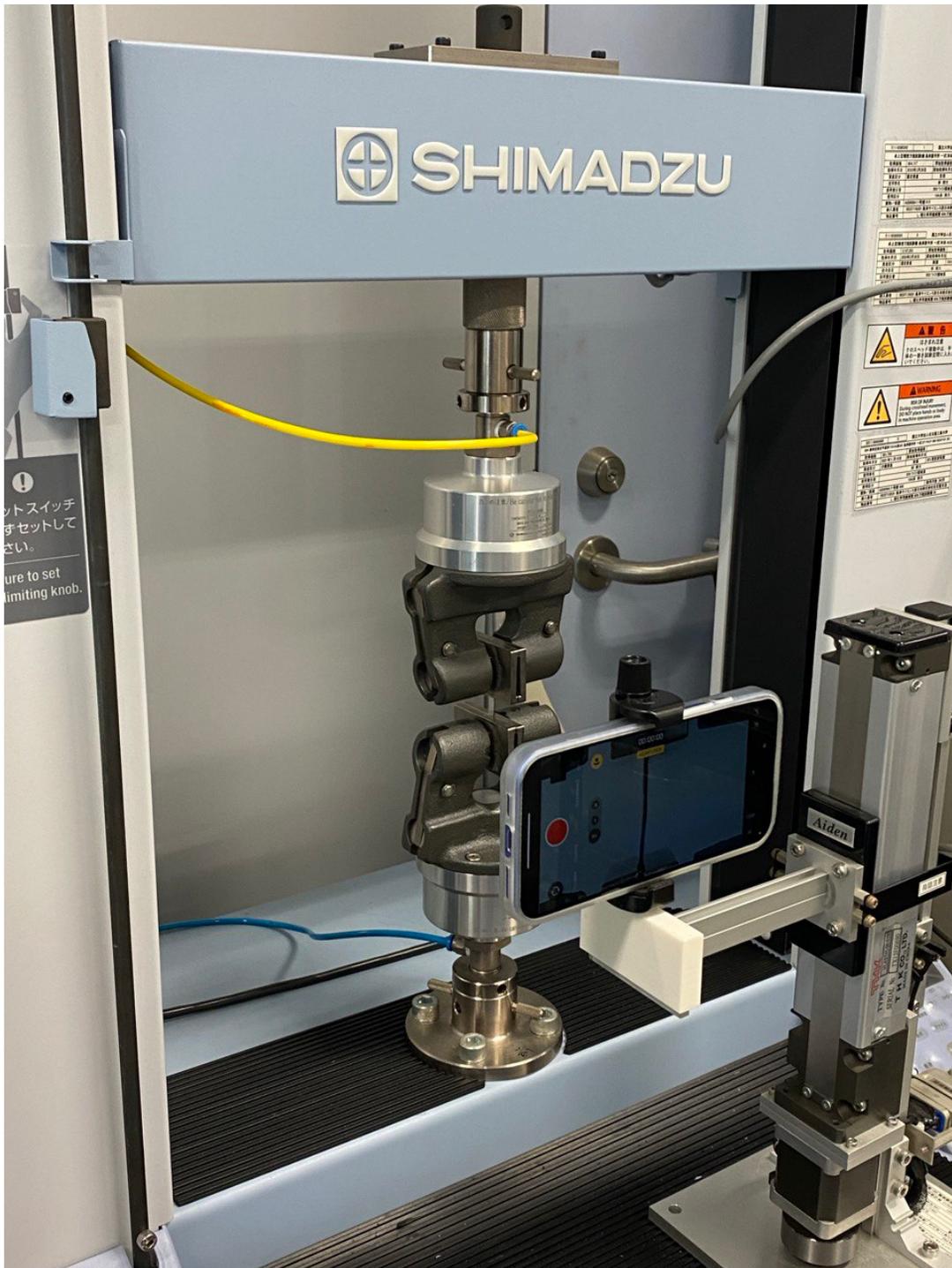

Figure S2. Experimental setup for the bending test
Image of the experimental setup for the bending test, showing the sample secured in the chucks of a universal testing machine. The sample is bent by moving it 5.5 mm closer at a speed of 0.1 mm/sec, with a dip coater used to lower the camera at half the stroke speed for recording.

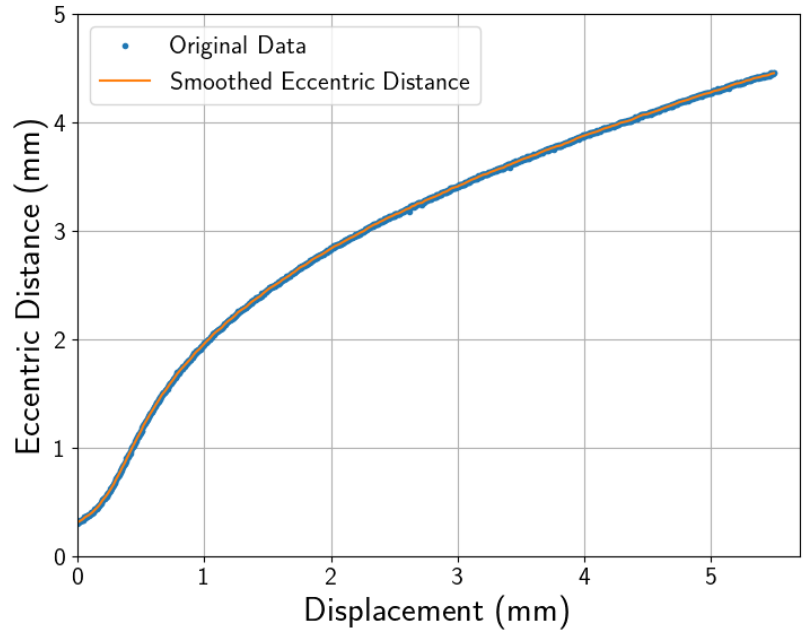

Figure S3. Smoothing of eccentric distance data
Smoothing of eccentric distance data using a Gaussian filter with a standard deviation of 3 for the Gaussian kernel.

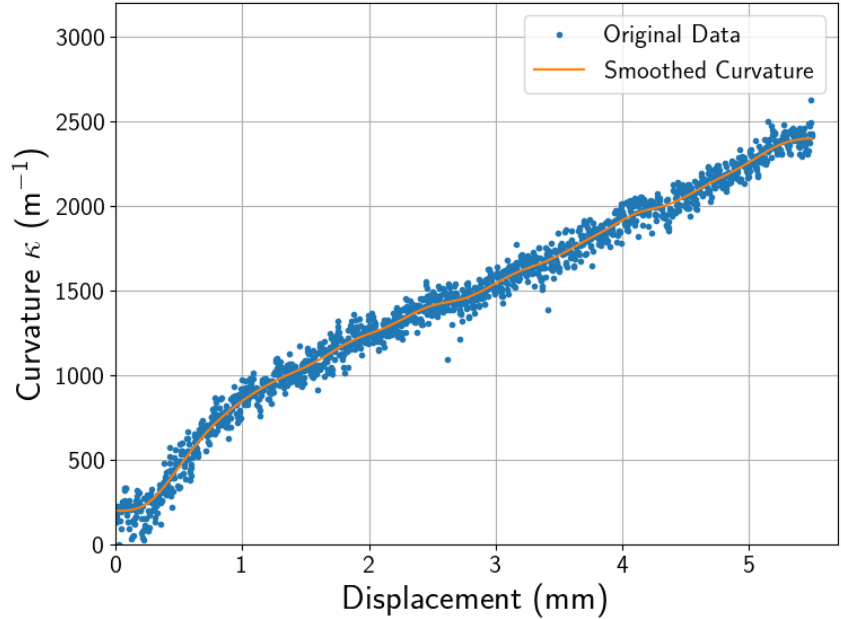

Figure S4. Smoothing of curvature data
Smoothing of curvature data using a Gaussian filter with a standard deviation of 40 for the Gaussian kernel.

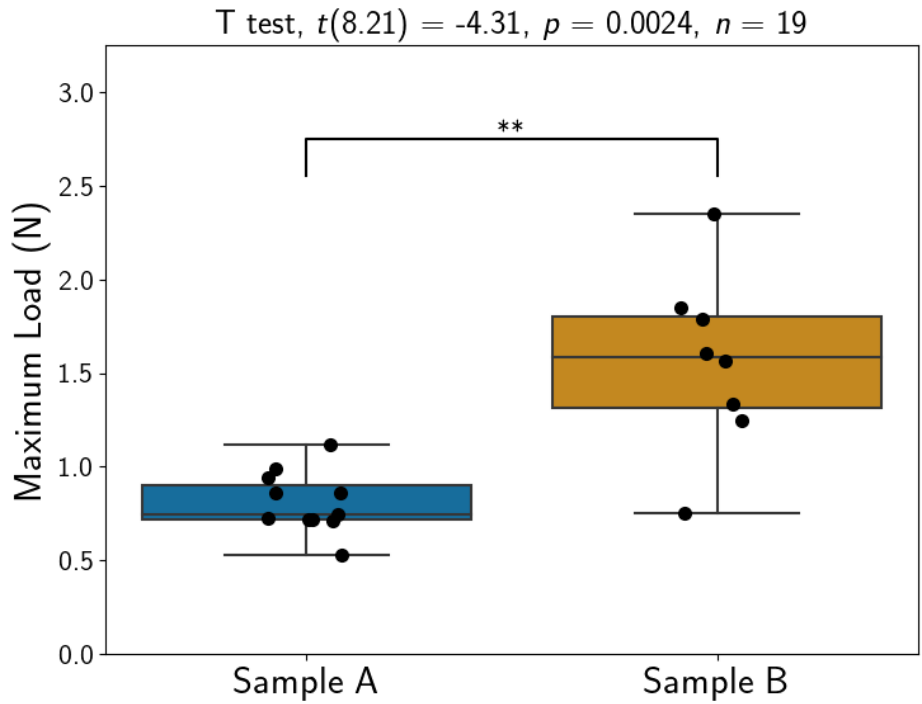

Figure S5. Comparison of maximum load in loading-unloading test
Box plot comparing the maximum load of samples A and B in the loading-unloading test. Welch's t-test was used for statistical analysis, with ** indicating $P < 0.01$.

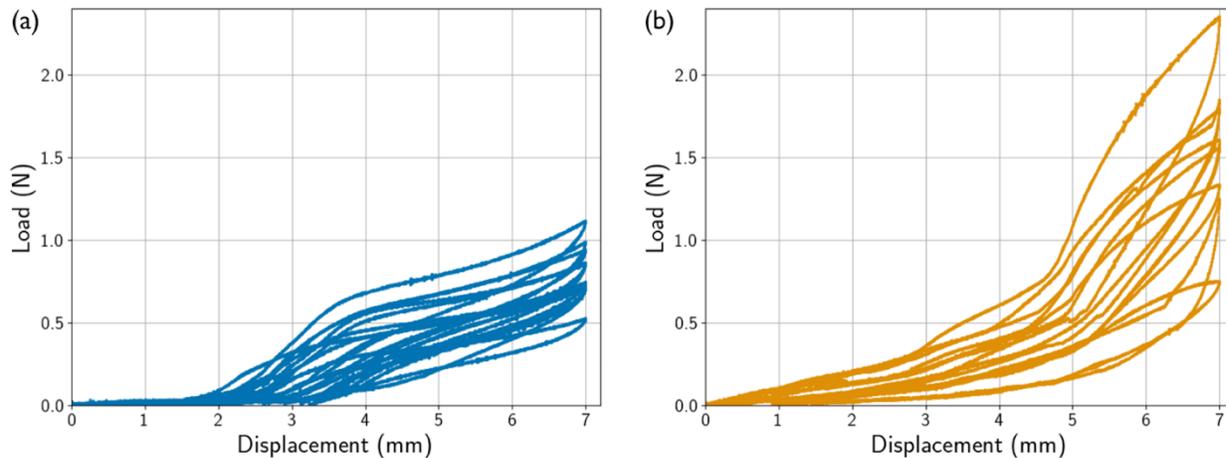

Figure S6. Load-displacement curves for sample A and B in the loading-unloading test
Load-displacement curves of sample A (a) and sample B (b) in the loading-unloading tests. Results from 11 samples of sample A and 8 samples of sample B are shown, confirming reproducibility.

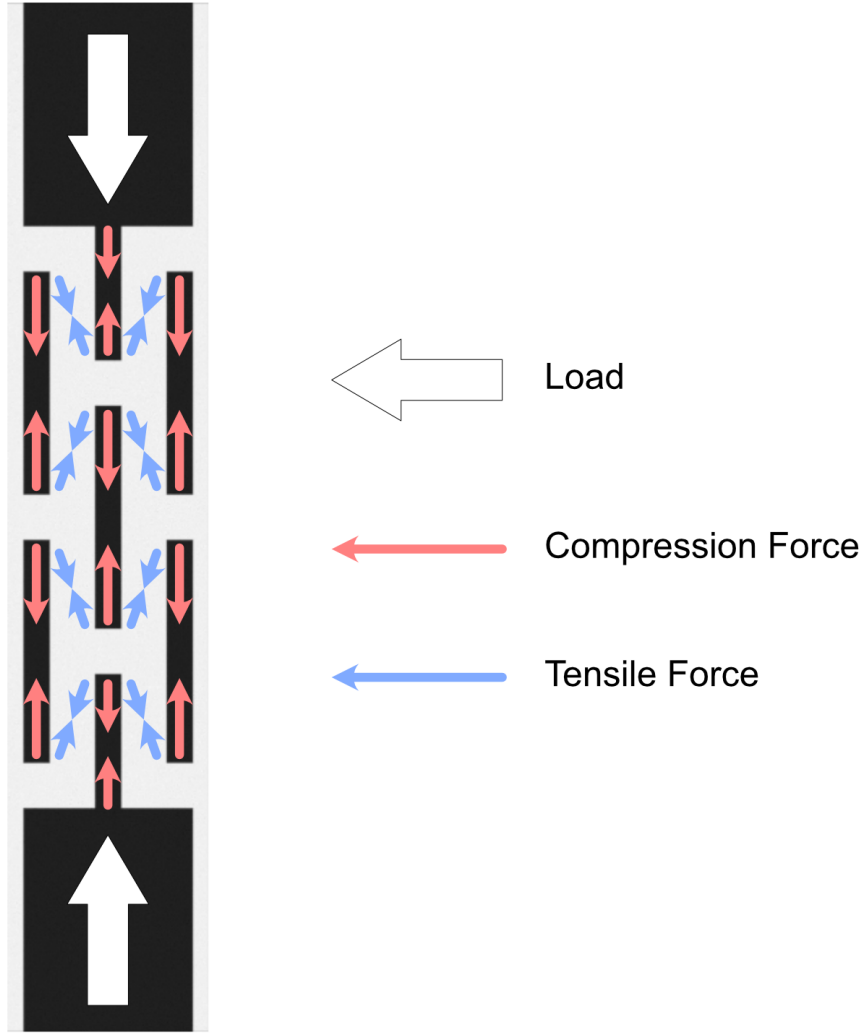

Figure S7. Schematic of sample B under vertical compression up to 0.5 mm displacement in the bending test
Schematic representation of sample B during vertical compression up to approximately 0.5 mm displacement, showing the adjacent rod tips moving apart, which causes membrane stretching and generates a restoring force.

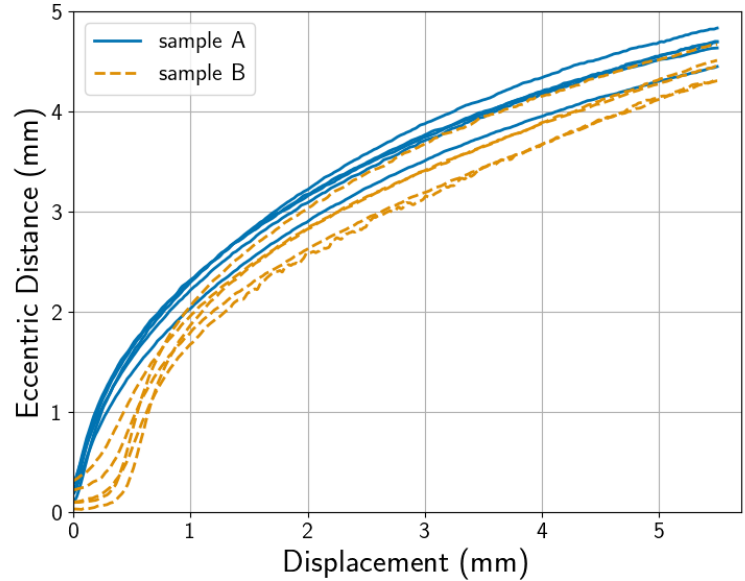

Figure S8. Eccentric distance–displacement curves for samples A and B in the bending test
Eccentric distance–displacement curves showing the monotonic increase in eccentric distance for samples A and B. In the initial compression phase, unlike sample A, there was no notable increase in eccentric distance for sample B. Resuls from five samples each of sample A and sample B are shown, confirming reproducibility.

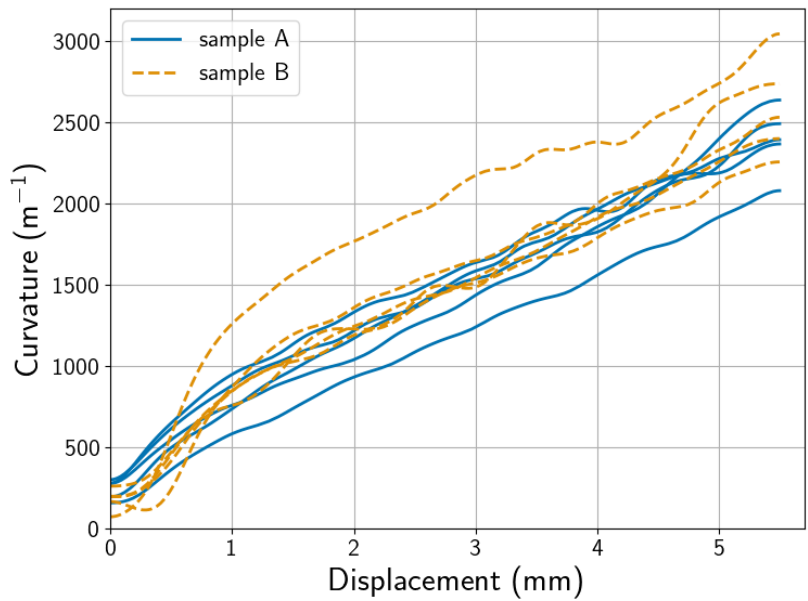

Figure S9. Curvature–displacement curves for samples A and B in the bending test
Curvature–displacement curves showing the monotonic increase in curvature for samples A and B. In the initial compression phase, unlike sample A, there was no notable increase in curvature for sample B. Results from five samples each of sample A and sample B are shown, confirming reproducibility.

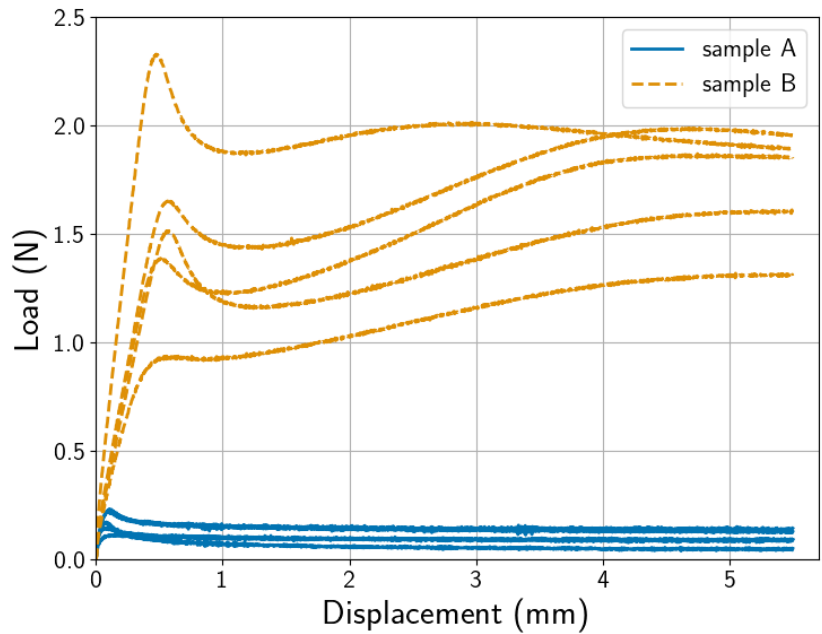

Figure S10. Load–displacement curves for samples A and B in the bending test
Load–displacement curves for samples A and B, showing a gradual increase in load for sample B as the displacement reaches approximately 1.5 to 4 mm. Results from five samples each of sample A and sample B are shown, confirming reproducibility.

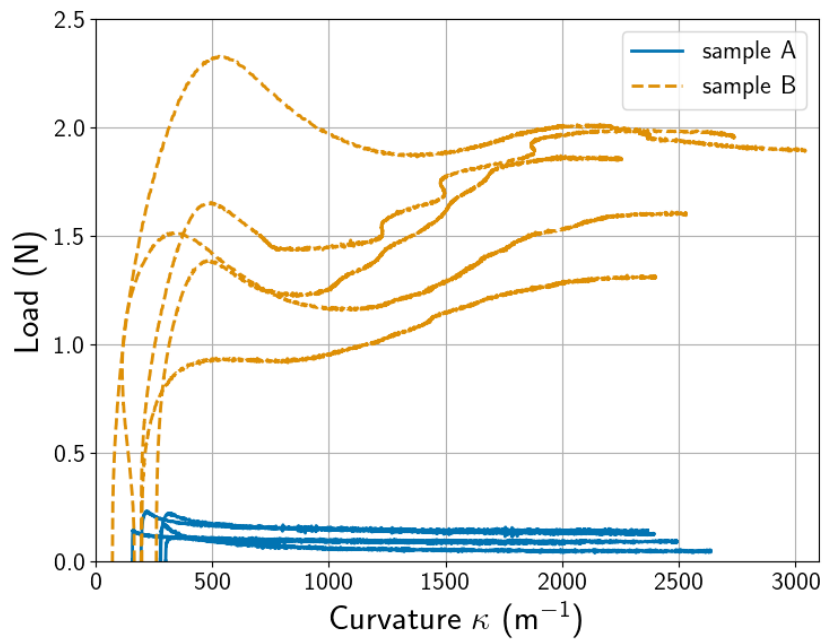

Figure S11. Load–curvature curves for samples A and B in the bending test
Curvature–load curves for samples A and B, illustrating the gradual load increase for sample B as the curvature ranges from approximately 1000 to 2000 m$^{-1}$. Results from five samples each of sample A and sample B are shown, confirming reproducibility.

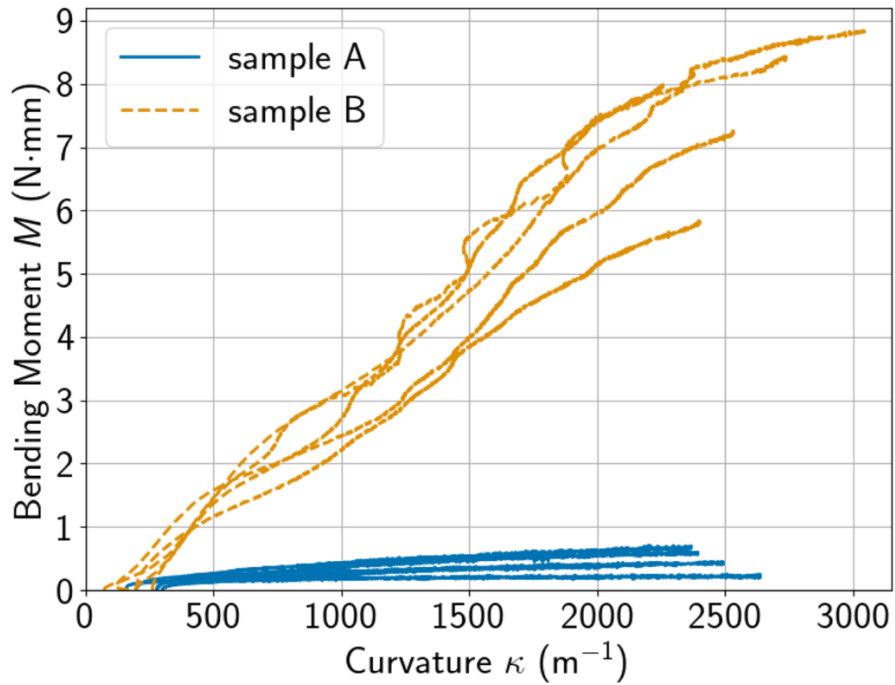

Figure S12. Moment–curvature curves for samples A and B in the bending test. Results from five samples each of sample A and sample B are shown, confirming reproducibility.

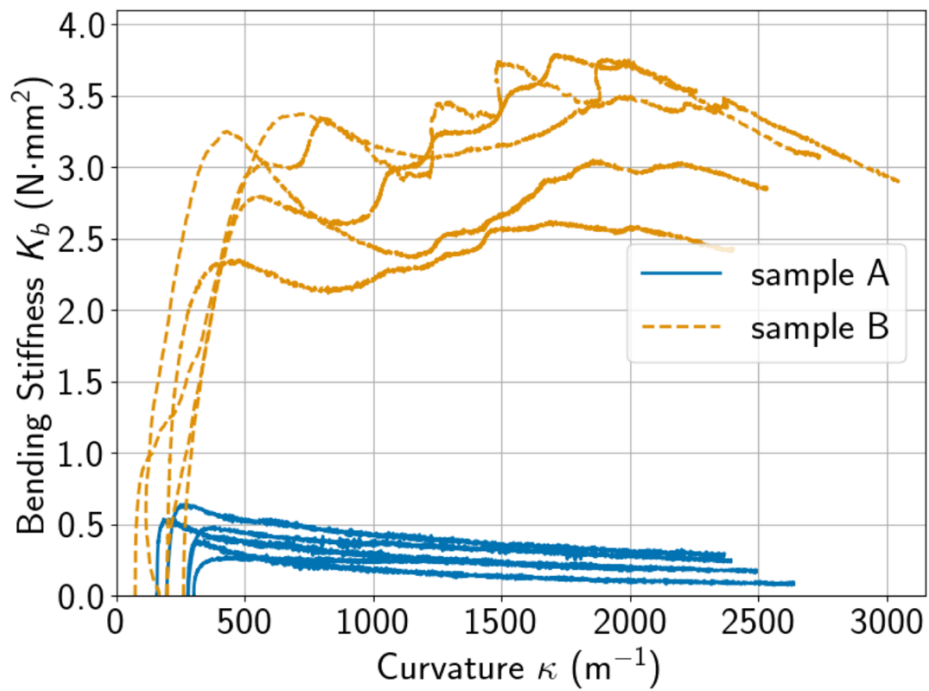

Figure 13. Stiffness–curvature curves for samples A and B in the bending test
Results from five samples each of sample A and sample B are shown, confirming reproducibility.